\documentclass[aps,pra,superscriptaddress,amsmath,amssymb,preprintnumbers,floatfix,showpacs,12pt]{revtex4}
\usepackage{amssymb}
\usepackage{epsfig}
\usepackage{subfigure}
\usepackage{graphicx}
\begin{document}
\title{Generic measure for the  quantum correlation of the two-qubit systems:
the average of the spin-correlation  elements  }

\author{Faisal A. A. El-Orany }
\email{el_orany@hotmail.com, Tel:006-2622876, Fax:0060386579404}
\affiliation{ Department of Mathematics and Computer Science,
Faculty of Science, Suez Canal University, Ismailia, Egypt; }
\affiliation{ Mathematical Modelling Lab, MIMOS Berhad, Technology
Park Malaysia, 57000 Kuala Lumpur, Malaysia}

\begin{abstract}
Based on the Pauli spin operators we develop the notion of the
spin-correlation matrix for the two-qubit system. If this matrix
is non-zero, the measure of the correlation between the qubits is
the average of the non-zero elements. Trivially, for zero matrix
the bipartite  is uncorrelated. This criterion turns out to be a
necessary and sufficient condition for the full correlation, where
 it includes information on both  entanglement and  correlation other
than entanglement. Moreover, we discuss to what extent this
criterion can give information on the entanglement of the system.
The criterion is generic in the sense that it can be applied to
mixed and  pure systems. Also, it can be easily extended to treat
the correlation of  multipartite systems. We compare the results
obtained from this criterion to those from concurrence for various
examples and we gain agreement regarding entanglement.
   We believe that this
criterion may have a wide range of potential applications in
quantum information theory.
\end{abstract}
 \maketitle

      \vspace{1cm}

Quantum entanglement is  a useful physical resource for quantum
information processing. For instance, the predicted capabilities
of a quantum computer  rely crucially on  entanglement
\cite{kimb}, and a proposed quantum cryptographic scheme achieves
security by converting shared entanglement into a shared secret
key \cite{[3]}. For both theoretical and potentially practical
reasons, it has become essential  to quantify  entanglement in
physical systems. Thereby, this subject  has attracted  massive
 interest recently.
Over the years, a number of different criteria for quantifying
entanglement of  physical systems  have been developed such as
negativity of the partial transpose \cite{peres},  concurrence
\cite{concr2,concr1},   correlation functions
\cite{palum1,palum2,palum3} and   quantum discord
\cite{wer1,wer2}. However, each one of these criteria has
advantages and disadvantages.

In this Letter, we propose a simple and novel correlation
criterion for the two-qubit system. The criterion  includes
information on both  entanglement and  correlation other than
entanglement. Throughout this letter, for  the entanglement we use
the phrase "nonclassical correlation", however, for the other
correlation we metaphorically use "classical correlation". The
criterion is generic in the sense that it works for both mixed and
pure systems. Also, we argue how one can extend it to assess the
correlation of the multipartite system. Moreover, we discuss to
what extent the information given by this criterion represents the
entanglement of the system.
 The steps for getting the criterion  can be
described as follows. We construct the spin-correlation matrix by
calculating  the correlation elements  of  all  pairs of the Pauli
spin operators of the qubits ( i.e. $\hat{\sigma}_x^{(j)},
\hat{\sigma}_y^{(j)}, \hat{\sigma}_z^{(j)}$).   If this matrix is
non-zero,  the proposed criterion, which can give  complete
information on the correlation of the bipartite, is the average of
the non-zero elements. Nonetheless, if all the elements of the
matrix are zero, the average will be automatically  zero declaring
uncorrelation. We call this criterion the correlation indicator
and denote it by $I$. It is worth mentioning that, in some
contexts, the concept of the correlation function has already been
considered in the literature \cite{palum1,palum2,palum3}. We
illustrate that $I\neq 0$ is  a necessary and sufficient condition
for the bipartite exhibiting correlation. Needless to say that if
entanglement exists, it will be a subset of this correlation.
Furthermore, we compare the results obtained from our criterion
with those
 from the concurrence $C(\rho)$ for various examples. The
main objective of this comparison is to show that when $C(\rho)$
exhibits entanglement, $I$ provides  quite similar information.
Besides,  $I$ can include additional information on the classical
correlation. It is valuable to mention the recent discovery that
 correlations other than entanglement can be responsible for
the quantum computational efficiency of the deterministic quantum
computation with one pure qubit \cite{datta}. Furthermore, the
comparison allows us to know when
 the information given by $I$ represents the  entanglement.
Precisely, we have noted that when $I>1/3$ the bipartite is
entangled, however, for $I\leq 1/3$ the bipartite may be
correlated classically or nonclassically. The limit $1/3$ is
obtained from the comparison with the entanglement  of the Werner
state. As we proceed, we shed some light on the sudden-death of
entanglement phenomenon (SDE) \cite{horod,eber1,
eber2,eber3,yonac,yonac2} in terms of  $I$. The SDE means that the
entanglement in the system disappears   in a finite interaction
time, in contrast to the exponential decay of the excited states
(or coherence) of some systems. Despite the fact that the SDE
sounds mysterious,
 there is no dynamical distinction
between separable and entangled states, since  quantum states can,
in general, evolve back and forth across the boundary between
 separable and entangled states \cite{wer1,wer2}. We show that $I$
cannot detect the SDE. In this respect its behavior is quite
 similar to that of the quantum discord  \cite{wer1,wer2}.
 We will describe  all these findings  in detail below.

Our starting point is the generic  density matrix of the
bipartite, which in the standard basis $\{ |e_1,e_2\rangle,
|e_1,g_2\rangle, |g_1,e_2\rangle, |g_1,g_2\rangle \}$ has the form: 
\begin{equation}\label{cor2}
\hat{\rho}=\left(%
\begin{array}{cccc}
 b_{11} & b_{12} & b_{13}& b_{14} \\
 b_{21} & b_{22} & b_{23}& b_{24}\\
  b_{31} & b_{32} & b_{33}& b_{34} \\
  b_{41} & b_{42} & b_{43}& b_{44} \\
\end{array}%
\right),
\end{equation}
where $b_{11}+b_{22}+b_{33}+b_{44}=1$.

Our main concern in this Letter is the evaluation of the full
correlation
 inherited in the
quantum system. Based on that  a state of a multipartite quantum
system is called correlated  if it cannot be factorized into
individual states. Mathematically, the quantum system described by
$\hat{\rho}$ is correlated  iff
it cannot be expressed as: 
\begin{equation}\label{ins1}
\hat{\rho}=\bigotimes_{j=1}^{n} \hat{\rho}_j,
\end{equation}
where $\hat{\rho}_j$ is the density matrix of the $j$th-party and
$n$ is the number of parties. The new criterion is working under
the condition that Eq. (\ref{ins1}) is satisfied regardless of
 the system   being  in the pure or in the mixed state.

 Next, we shed   some light   on the concurrence, which will be
 the benchmark for the suggested criterion regarding  entanglement.  The
concurrence $C(\rho)$  can be described as follows
\cite{concr2,concr1}: 
\begin{equation}\label{death1}
 C(\rho)={\rm
 max}\{0,\sqrt{\nu_1}-\sqrt{\nu_2}-\sqrt{\nu_3}-\sqrt{\nu_4}\},
\end{equation}
 where the
elements $\nu_j$ are the eigenvalues in decreasing order of the
auxiliary matrix
\begin{equation}\label{death2}
 \xi=\hat{\rho}(\hat{\sigma}_y\otimes\hat{\sigma}_y)\hat{\rho}^*(\hat{\sigma}_y\otimes\hat{\sigma}_y),
\end{equation}
 where $\hat{\rho}^*$ denotes the complex conjugation of $\hat{\rho}$ in the standard
basis and $\hat{\sigma}_y$ is the Pauli-spin matrix. We have
$1\geq C(\rho)\geq 0$, where $C(\rho) = 0$ indicates separability
(zero entanglement) and $C(\rho) = 1$ means maximal entanglement.

 Now, we are in a position to develop
 the new criterion, which quantifies  the correlation of the bipartite. For this purpose, we define
  the spin-correlation  matrix as:
\begin{equation}\label{cor1}
I_{xyz}=\left(%
\begin{array}{cccc}
 I_{xx} & I_{xy} & I_{xz} \\
 I_{yx} & I_{yy} & I_{yz} \\
 I_{zx} & I_{zy} & I_{zz} \\
\end{array}%
\right),
\end{equation}
where
\begin{equation}\label{gr2}
   I_{ij}=|\langle\hat{\sigma}_i^{(1)}\hat{\sigma}_j^{(2)}\rangle|
   -|\langle\hat{\sigma}_i^{(1)}\rangle\langle\hat{\sigma}_j^{(2)}\rangle|,\quad
   i,j=x,y,z.
\end{equation}
The modulus in (\ref{gr2}) is given to achieve a quantitative
description  for the correlation of the bipartite. In this regard,
the elements  $I_{ij}$
 are quite different from  the covariance ones given in
\cite{palum2}. The reason for using all  possible elements
$I_{ij}$ in the matrix (\ref{cor1}) is that the occurrence of
correlation (and consequently entanglement) at a time instant $t$
necessarily implies the existence of at least a couple of
operators $\hat{\sigma}_i^{(j')}$ acting on the bidimensional
Hilbert spaces of  spin $1$ and $2$ such that the corresponding
element
 is non-zero.
When $I_{ij}\neq 0, i\neq j$ or $I_{ii}\neq 0$ means that the
bipartite is correlated in the $ij$-plane or in the $i$-axis
direction. Actually, there is a similarity between the structure
of the matrix (\ref{cor1}) and that of the moment of inertia
tensor matrix of the 3 dimensional rigid body as well as that of
the covariance matrix of a trivariate random vector whose
probability density function is proportional to the pointwise
density of the rigid body. For any arbitrary quantum atomic state,
one can easily verify that
$|\langle\hat{\sigma}_i^{(1)}\hat{\sigma}_j^{(2)}\rangle|\geq
   |\langle\hat{\sigma}_i^{(1)}\rangle\langle\hat{\sigma}_j^{(2)}\rangle|$
   for all $i,j$. Also, the modulus of the expectation value of any
   multiples of the
   Pauli spin operators  always exists inside or on the surface of the Bloch sphere.
Thereby, we have  $0\leq I_{ij}\leq 1$, where $I_{ij}=0\quad (1)$
means that the bipartite is disentangled (maximally
correlated/entangled) in the $ij$-plane. Moving forward, we show
that the matrix (\ref{cor1}) includes all necessary information
about the correlation/entanglement of the bipartite. The elements
of the density matrix (\ref{cor2}) can be categorized  into three
groups, namely, diagonal, anti-diagonal $\{b_{14}, b_{23}, b_{32},
b_{41}\}$, and the remaining elements. The first group contributes
to  the element $I_{zz}$, the second one gives information on
$\langle\hat{\sigma}_i^{(1)}\hat{\sigma}_i^{(2)}\rangle$ with
$i=x,y$, while  the last group provides  information on
$\langle\hat{\sigma}_i^{(1)}\hat{\sigma}_j^{(2)}\rangle$ and
$\langle\hat{\sigma}_l^{(k)}\rangle$ where $i,j=x,y,z, i\neq
j,\quad  l=x,y$ and $k=1, 2$.  In order  to obtain full
information on the correlation of the bipartite, one has to deal
with all elements in the matrix (\ref{cor1}), not only some  of
them. This contrasts with the early study in \cite{palum2}, where
the authors confined themselves  to two elements only. We
elaborate this point  by giving the following example. Assume that
we would like to assess the correlation of the  of the
quasi-Bell-state:
\begin{equation}\label{gr3}
    |\Psi\rangle=\cos \theta |e_1,e_2\rangle+ \sin\theta
    |g_1,g_2\rangle.
\end{equation}
 If we  restrict ourselves  to the elements $I_{i\neq j} $,
 we will find that the bipartite is disentangled, which is not
true. Considering all these facts, the new  criterion (, i.e. the
correlation indicator $I$), which can  give  accurate  information
on the correlation
(classical or nonclassical)  of the bipartite, is: 
\begin{equation}\label{cor3}
I=\frac{1}{N}\sum\limits_{i,j=x,y,z} I_{ij},
\end{equation}
where $N$ represents the number of the non-zero elements in
$I_{xyz}$. As we mentioned above, if   $I_{ij}=0$ for all $i$ and
$  j$, then  $I=0$ and the bipartite is disentangled.
 The indicator   $I$ ranges between $0$ (disentangled) and
$1$ (maximally correlated/entangled) bipartite. It is obvious that
$I$ has various merits. It is a well-behaved quantity and can be
applied for both  pure and  mixed systems.
 It is easy to be
manipulated since it deals with the moments of the spin operators.
It  gives   a simple  means  to  interpret and to measure the
correlation.  Furthermore, the criterion (\ref{cor3}) can be
easily extended to treat  the correlation/entanglement of the
multipartite systems.

Our next task is to check the efficiency of  the criterion
(\ref{cor3}). This may be done by applying it to some examples and
comparing the obtained results with those from the concurrence. To
achieve this goal we give   five examples; two of them (examples 4
and 5) are devoted to the SDE.

\noindent {\bf Example (1)}:
 Assume we would like to
quantify  the correlation between  two qubits of the
 three-particle
Greenberger-Horne-Zeilinger state (GHZ) \cite{Greenberger}, which
has  the form, e.g.,:
\begin{equation}\label{gr}
    |\Psi_+\rangle=\frac{1}{\sqrt{2}}[|e_1,e_2,e_3\rangle+|g_1,g_2,g_3\rangle].
\end{equation}
The density matrix of   any  two qubits of the state (\ref{gr}),
say $1,2$, can be evaluated by tracing over the third particle as:
\begin{equation}\label{gr1}
    \hat{\rho}=\frac{1}{2}[|e_1,e_2\rangle\langle e_2,e_1|+|g_1,g_2\rangle\langle
    g_2,g_1|].
\end{equation}
Foremost, the density matrix (\ref{gr1}) represents a maximally
mixed state. In terms of the concurrence,  state (\ref{gr1}) is
disentangled, where $C(\rho)=0$. This is related to the definition
of  $C(\rho)$, which is based on the spin flip transformation.
Nevertheless,  state (\ref{gr1}) is maximally correlated  with
respect to our criterion, where $I=1$.

\noindent {\bf Example (2)}:
 We study the entanglement of  the  quasi-Bell-state
(\ref{gr3}). For this state,  the indicator $I$ can be expressed as: 
\begin{equation}\label{cor5}
I=\frac{1}{3}(
I_{xx}+I_{yy}+I_{zz})=\frac{1}{3}[\sin^2(2\theta)+2|\sin(2\theta)|].
\end{equation}
Also, its   concurrence  takes the form:
\begin{equation}\label{gr5}
  C(\rho)= \sin^2(2\theta).
\end{equation}
It is evident from  equations (\ref{cor5}) and (\ref{gr5}) that
both  $I$ and $C(\rho)$ have almost the same  behavior. In this
case, $I$ gives information on the entanglement.

{\bf Example (3)}. Here we shed some light on correlation of  the
general case of $2$-q-bit pairs. Also we compare the  results
obtained from our criterion with those given in \cite{eng1,eng2}
as well as with the concurrence. To this end, we use the general
form of a $2$-q-bit state   \cite{eng1,eng2}:
\begin{equation}\label{gr8new}
\widetilde{\hat{ \rho}}=
\frac{1}{4}[1+\overrightarrow{\sigma}.s^{\downarrow}
+\overrightarrow{t}.\tau^{\downarrow}+\overrightarrow{\sigma}.
\overrightarrow{^{\downarrow} C}.\tau^{\downarrow}],
\end{equation}
where $\overrightarrow{^{\downarrow} C}=\langle
\sigma^{\downarrow} \overrightarrow{\tau} \rangle$. We employ the
terminology and the notational conventions of \cite{eng1,eng2}, in
which  the individual q-bits  are described by means of  analogs
of Pauli's spin vector operator: $\overrightarrow{\sigma}$ for the
first q-bit, $ \overrightarrow{\tau}$ for the second. The arrows
$\rightarrow, \downarrow, \overrightarrow{^{\downarrow} } $ denote
row vector, column vector and the cross dyadic. In
\cite{eng1,eng2} it has been shown that the characterization of
the $2$-q-bit states produced by some source requires the
experimental determination of $15$ real parameters. Based on the
knowledge of these parameters,  the $2$-q-bit state has been
classified into six classes of families of locally equivalent
states. Simple criteria have been stated  for checking a given
state's positivity and separability through fulfilling certain
inequalities. The technique has been employed in finding the
 Lewenstein-Sanpera decompositions (LS) \cite{eng3}  for the state $\widetilde{\hat{\rho}}$.
  This will be our main concern in this comparison.
   Precisely, LS states that  any $2$-q-bit state $\widetilde{\hat{\rho}}$ can be
written as a mixture of a separable state $\hat{\rho}_{sep}$ and a
non-separable pure state $\hat{\rho}_{pure}$:
\begin{equation}\label{gr10new}
\widetilde{\hat{ \rho}}= \lambda \hat{\rho}_{sep}+(\lambda-1)
\hat{\rho}_{pure}.
\end{equation}
There are many different such LS decompositions with varying
values of $\lambda$. Among them is the unique optimal
decomposition, the one with the largest $\lambda$ value,
\begin{equation}\label{gr11new}
\widetilde{\hat{ \rho}}= \mu \hat{\rho}^{(opt)}_{sep}+(1-\mu)
\hat{\rho}^{(opt)}_{pure}, \quad  \mu={\rm max}\{\lambda\}.
\end{equation}
 In this case $\mu$ represents
the degree of separability of $\widetilde{\hat{ \rho}}$; the value
$\mu = 0$ obtains only if $\widetilde{\hat{ \rho}}$ itself is a
nonseparable pure state.
 For the separability of (\ref{gr8new}), we restrict ourselves
to two examples, which have been studied in detail in
\cite{eng1,eng2}, namely, Werner states and the rank-2 states.

For Werner states, the generic density matrix  (\ref{gr8new})
takes the form:
\begin{equation}\label{gr9new}
\hat{ \rho}_w= (1-x)\hat{ \rho}_{chaos}+x\hat{ \rho}_{Bell}=
\frac{1}{4}[1-x\overrightarrow{\sigma}.\sum\limits_{k=1}^{3}
e^{\downarrow}_k \overrightarrow{n}_k. \tau^{\downarrow}],
\end{equation}
where the subscripts $choas$ and $Bell$ stand for the $2$-qubit
density matrices of the thermal and the Bell state, respectively.
For the sake of clarification,  in the transition from
(\ref{gr8new}) to (\ref{gr9new}) one has $s^{\downarrow}=0,\quad
\overrightarrow{t}=0, \quad \overrightarrow{^{\downarrow} C}=
-x\sum\limits_{k=1}^{3} e^{\downarrow}_k \overrightarrow{n}_k $.
 For this state,
the degree of separability $\mu$  \cite{eng1}, concurrence
$C(\rho)$  and correlation indicator $I$ for (\ref{gr9new}) can be
expressed as:
\begin{equation}
\mu  = \left\{
\begin{array}{lr}
1
\;\;&{\rm if}\;-\frac{1}{3}\leq x\leq \frac{1}{3}  ,\\
\frac{3}{2}(1-x) \;\;&{\rm if}\;\frac{1}{3}< x\leq 1,
\end{array}
\right. \quad C(\rho)= {\rm max}\{0,\frac{3}{2}(x-\frac{1}{3})\},
\quad \quad I=|x|. \label{scf4}
\end{equation}
The result of $\mu$  agrees well with both the numerical findings
of Lewenstein and Sanpera \cite{eng3} and the concurrence.
 Moreover, it is clear that $\mu, C(\rho)$ and $I$  agree
with each other for the extreme values. It should be borne in mind
 that $I$ has been derived based on the definition (\ref{ins1}).
Based on that,  the correlation found in the interval
$-\frac{1}{3}\leq x\leq \frac{1}{3}$ in $I$ can be understood
classically. Inspired by this observation we can modify $I$, for
this case, to provide the nonclassical correlation in
$\frac{1}{3}< x\leq 1$ by subtracting $1/3$, i.e. $I-1/3$.
However, the modified version needs to be normalized to  unity at
$x=1$, as it should be. Then we get $\widetilde{I} =C(\rho)$. It
is evident that the degree of separability $\mu$ and
$\widetilde{I}$ give identical information on the entanglement of
$\hat{ \rho}_w$. Also, one can conclude: for any bipartite when
$I>1/3$ the correlation can be interpreted as entanglement.

Now we give attention to  states of rank $2$ \cite{eng2}:
\begin{equation}\label{gr19new}
\hat{ \rho}_{rank2}= \frac{1}{4}[1+(\sigma_3+x\tau_3)\sin(\theta)+
(\sigma_1\tau_1-x\sigma_2\tau_2)\cos(\theta)+x\sigma_3\tau_3],
\end{equation}
with $-1<x<1$.  The generic form for states of rank 2 can be found
in \cite{eng1}.  The degree of separability, concurrence and
correlation indicator $I$ for (\ref{gr19new}) are:
\begin{eqnarray}\label{ebr51new}
\begin{array}{lr}
\mu=\left\{
\begin{array}{lr}
1
\;\;&{\rm if}\;\cos(\theta)=0   ,\\
1-|x| \;\;&{\rm if}\;\cos(\theta)\neq 0,
\end{array}
\right.\quad\quad
C(\rho)=|x\cos(\theta)|,\\
I=\frac{1}{3}[(1+|x|)|\cos(\theta)|+|x|\cos^2(\theta)].
\end{array}
\end{eqnarray}
It is obvious that these  three criteria  provide the same
information for the cases $\cos(\theta)=0$ and $
x=\cos(\theta)=1$. However, for $x=0, \cos (\theta)=1$, the
bipartite is separable in terms of $\mu$ and $C(\rho)$, but
$I=1/3$. This confirms  the above observation: the correlation
exhibited by $I$ is nonclassical whenever $I
>1/3$.
 For certain range of the  parameters, $C(\rho)$ and
$ I$ have quite  similar behavior, but this is not the case for
$\mu$. This is noticeable by comparing  different expressions in
(\ref{ebr51new}). As an example, when $x=1, \cos(\theta)=1/2$ we
have $\mu=0 $ (maximum entanglement), whereas $C(\rho)=1/2$ and
$I=5/12$.

\noindent {\bf Example (4)}: This example and the following one
have been very often used in the literature to establish the
concept of the SDE \cite{horod,eber1, eber2,eber3,yonac,yonac2}.
Thus, we use them to discuss the SDE in the framework of $I$. We
consider the evolution of the state (\ref{gr3}) with the two-mode
Jaynes-Cummings model \cite{eber1,eber2,eber3,yonac,yonac2}, which
may be described by the interaction Hamiltonian:
\begin{equation}\label{ebr1}
\hat{H}_I= \lambda \sum\limits_{j=1}^{2}(\sigma _{+}^{(j)}%
\hat{a}_j+\sigma _{-}^{(j)}\hat{a}^{\dagger }_j),
\end{equation}
where $\sigma _{+}^{(j)}\quad (\hat{a}^{\dagger }_j) $ and $\sigma
_{-}^{(j)} \quad (\hat{a}_j)$ are the raising and the lowering
atomic (field) operator of the $j$th-party, respectively,
$\lambda$ is the coupling constant and we take $\hbar=1$. For the
initial field state $|0_1,0_2\rangle$, the elements of the matrix
(\ref{cor2}) of the two qubits take the forms \cite{yonac}:
$b_{11}=\cos^2\theta \cos^4T,\quad
b_{22}=b_{33}=\frac{1}{4}\cos^2\theta \sin^2(2T), \quad
b_{44}=\cos^2\theta \sin^4T+\sin^2\theta,\quad
b_{14}=b_{41}=\frac{1}{2}\sin(2\theta) \cos^2T$, with
$T=t\lambda$, and the other elements are zero. The concurrence and
the indicator  $I$ of the two qubits of this system read:
\begin{equation}\label{ebr2}
C(\rho)= 2\cos^2T\cos^2\theta \quad {\rm max}\{0,|\tan
\theta|-\sin^2 T\},
\end{equation}
\begin{eqnarray}\label{ebr3}
\begin{array}{lr}
I=\frac{1}{3} \{
1-\sin^2(2T)\cos^2\theta-[\cos(2T)\cos^2\theta-\sin^2\theta]^2
\\
+|\sin(2\theta)\cos^2T+\frac{1}{2}\cos^2\theta\sin^2(2T)|
+|\sin(2\theta)\cos^2T-\frac{1}{2}\cos^2\theta\sin^2(2T)| \}.
\end{array}
\end{eqnarray}
Comparison between (\ref{ebr2}) and (\ref{ebr3}) reveals that
$C(\rho)$  exhibits SDE only when $|\tan \theta|\leq\sin^2 T$,
while   $I$ cannot. The reason is that $I$ includes the classical
and the nonclassical correlation, while $C(\rho)$ has the
nonclassical one. In Figs. 1(a) and (b) we depict $I$ and
$C(\rho)$ for two different cases $\theta=\pi/4$ and $\pi/6$,
respectively. From Fig. 1(a), the behavior of the two criteria  is
almost the same, where, for this case, $C(\rho)=\cos^4T$ and
$I=\frac{1}{3}(2+\cos^2T)\cos^2T$. Fig. 1(b) shows that $I$ has
smooth oscillatory behavior without SDE, whereas $C(\rho)$
non-smoothly becomes and stays zero for a finite interval of time.
As a result  of the  lossless nature of the evolution,  the
original entanglement value  occurs in a periodic way following
each sudden death event. The beauty of this example is that SDE
occurs without decoherence of the traditional type.
 The origin of
the occurrence of SDE in  $C(\rho)$ is in the cut-off existing in
the definition of the concurrence (\ref{death1}).
 As we proceed, one can observe that $I$
vanishes only at discrete instants, i.e. $T=\pi/2$, declaring the
disentanglement in the bipartite. One can verify this result by
substituting $(T,\theta)=(\pi/2,\pi/6)$ into the density matrix of
the bipartite, which reduces to $\hat{\rho}=|g_1\rangle\langle
g_1|\bigotimes |g_2\rangle\langle g_2|$, i.e. the bipartite
becomes pure and separable. In other words,   $I=0$ means that the
information is locally accessible and can be obtained by distant
independent observers without perturbing the bipartite state. The
horizontal dashed line in Fig. 1(b) confirms our observation: for
$I> 1/3$ the bipartite is entangled.

\begin{figure}[h]%
 \centering
 \subfigure[]{\includegraphics[width=6cm]{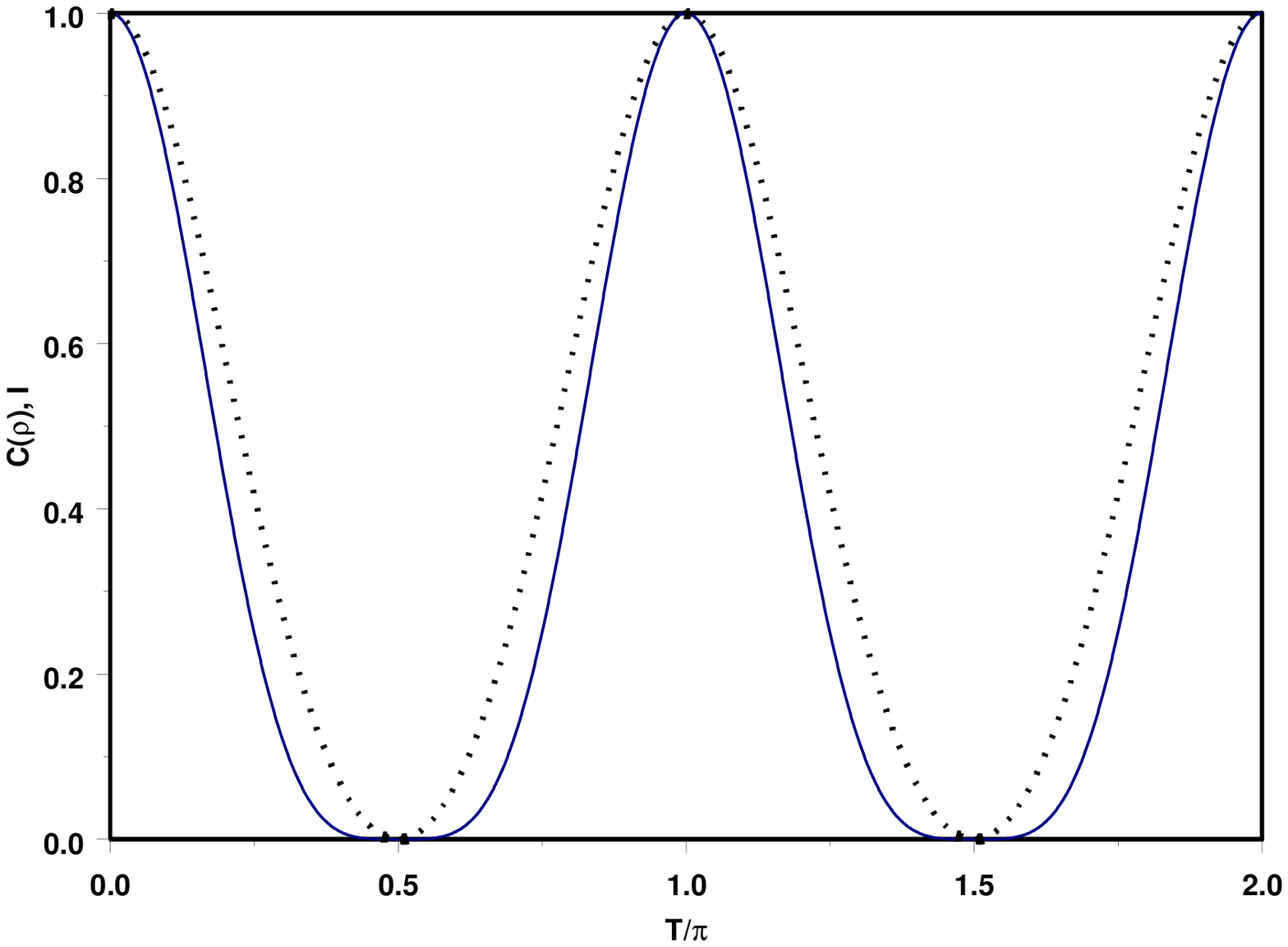}}
\subfigure[]{\includegraphics[width=6cm]{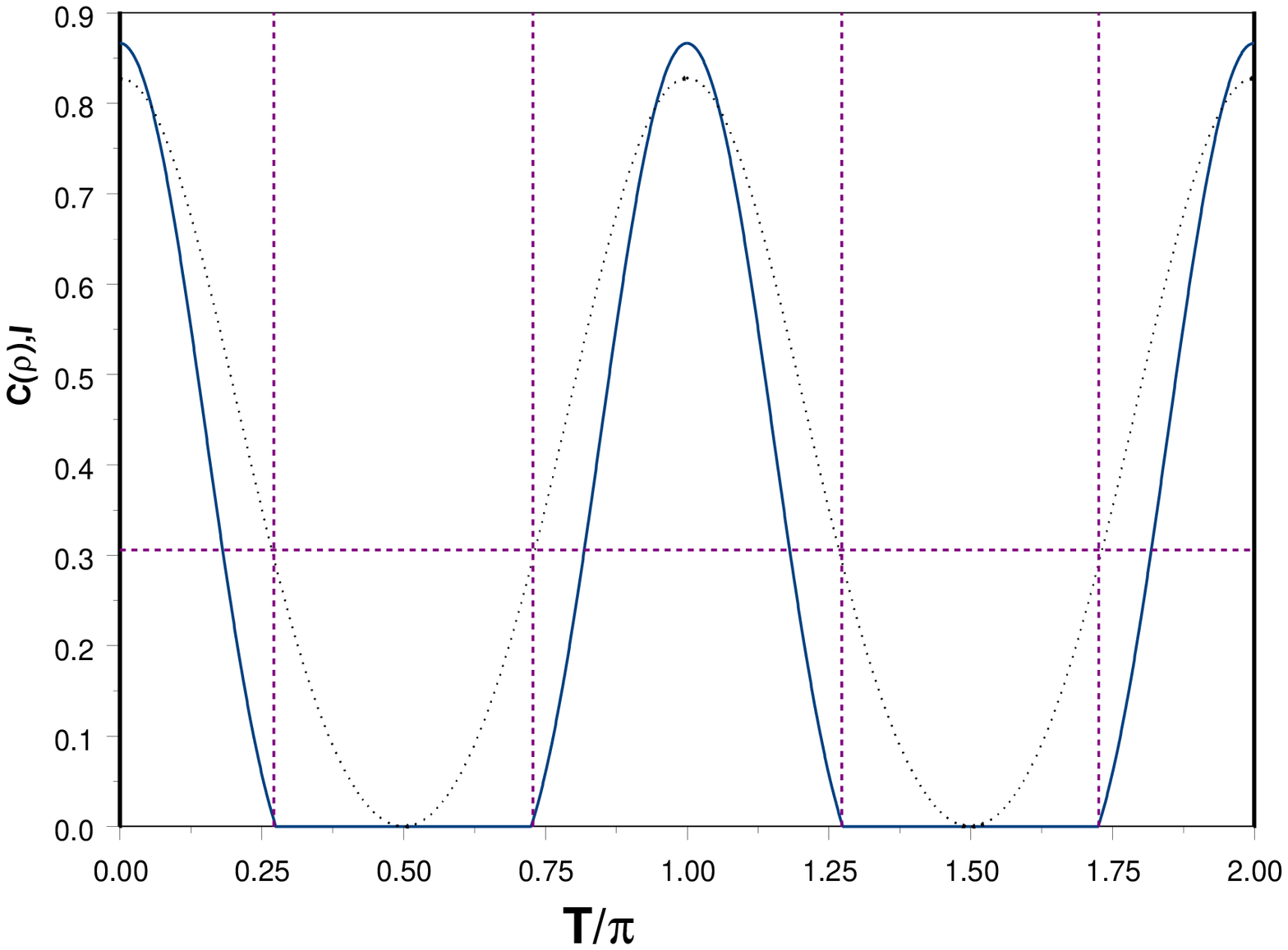}}
    \caption{Atom-atom entanglement  quantified by $C(\rho)$ (solid curve) and
   $I$ (dashed curve)  when the atoms are initially prepared in  state (\ref{gr3})
with $\theta=\pi/4$ (a)  and $\pi/6$ (b). The vertical and
horizontal dashed lines in (b) to show the interval of the
classical  correlation in $I$ or the the segment of  SDE in
concurrence.
 The time scale is the vacuum Rabi period.
  }
 \label{fig1}
\end{figure}

\noindent {\bf Example (5)}: Here, we consider two two-level atoms
coupled individually to two cavities which are initially in their
vacuum states. In the framework of system-plus-environment, the
two two-level atoms are identified as the system of interest,
whereas the two cavities serve as the environments. The total
Hamiltonian describing this scenario is \cite{eber1}:
\begin{eqnarray}\label{ebr4new}
\begin{array}{lr}
\hat{H}=\frac{\omega_1}{2}\sigma
_{z}^{(1)}+\frac{\omega_2}{2}\sigma _{z}^{(2)}+\sum_{k}
\left[w^{(1)}_k\hat{a}^{\dagger}_k\hat{a}_k+w^{(2)}_k\hat{b}^{\dagger}_k\hat{b}_k\right]\\
\\
+ \sum_{k} \left\{\left[g_k \sigma _{+}^{(1)}\hat{a}_k+g^{*}_k
\sigma _{-}^{(1)}\hat{a}^{\dagger}_k\right] + \left[f_k \sigma
_{+}^{(2)}\hat{b}_k+f^{*}_k \sigma
_{-}^{(2)}\hat{b}^{\dagger}_k\right]\right\},
\end{array}
\end{eqnarray}
where $g_k, f_k$ are coupling constants and the other notations
have the same standard meaning.  When   the atoms are initially
entangled with each other but not with the cavities, the solution
of the system has been  derived already in \cite{eber1} via the
Born-Markov approximation. We focus our attention on the case that
the atoms are initially in the mixed entangled state with
$b_{11}=a/3 \geq 0, b_{22}= b_{33}= b_{23}=b_{32}=1/3,\quad
b_{44}=(1-a)/3$ and the other elements  are zeros. Therefore, the
dynamical density matrix elements of the two atoms have the form
\cite{eber1}:
\begin{eqnarray}\label{ebr5new}
\begin{array}{lr}
b_{11}(t)=\frac{1}{3}\gamma_1^2\gamma_2^2 a, \quad
b_{22}(t)=\frac{1}{3}( \gamma_1^2 +\gamma_1^2 \Omega_2^2 a), \quad
\quad b_{33}(t)= \frac{1}{3}(\gamma_2^2
+\gamma_2^2 \Omega_1^2 a),\\
\\
\quad b_{44}(t)=\frac{1}{3}(1-a+  \Omega_1^2+ \Omega_2^2+
\Omega_1^2 \Omega_2^2 a),\quad
b_{23}(t)=b_{32}(t)=\frac{1}{3}\gamma_1\gamma_2,
\end{array}
\end{eqnarray}
 and the other elements are zero. In using  the Markov limit results
and assuming  the cavities are similar, we have
$\gamma_1=\gamma_2= \exp(-\Gamma t/2),\quad
\Omega_1=\Omega_2=\sqrt{1-\exp(-\Gamma t)}$, where the $\Gamma$ is
 the Einstein A coefficient for the two-level atoms in the
cavities. The concurrence and the correlation indicator  of this
case are \cite{eber1}:
\begin{eqnarray}\label{gr5new}
\begin{array}{lr}
  C(\rho(T))= \textrm{max}\{0,
  |b_{23}(T)|-\sqrt{b_{11}(T)b_{44}(T)}\},\\
  I(T)=\frac{4}{3}b_{23}(T)+\frac{1}{3}[1-4b_{22}(T)]-\frac{1}{3}\left[1-2b_{11}(T)-2b_{22}(T)\right]^2,
\end{array}
\end{eqnarray}
where $T=\Gamma t$. For the initial state the expressions
(\ref{gr5new})  reduce to
\begin{equation}\label{gr6new}
  C(\rho(0))= \frac{2}{3}[1-\sqrt{a(1-a)}], \quad
I(0)=\frac{2}{27}(2a-4a^2+7).
\end{equation}
It is evident that the initial two qubits are always entangled in
terms of the two criteria (where $I>1/3$) regardless of the values
of $a$. In Fig. 2(a) and (b) we have plotted $ C(\rho(T))$ and
$I(T)$, respectively.   Fig. 2(a) presents the well-known figure
of the SDE. This figure shows that within the general exponential
character evident in (\ref{ebr5new}), disentanglement can be
completed, i.e. $C(\rho(T))=0$, in a finite time while the local
decoherences need an infinite time. For instance, when $a=1$, this
finite time is $T=\ln (\frac{2+\sqrt{2}}{2})$. The comparison
between figure 2(a) and 2(b) reveals when  $C(\rho(T))$ exhibits
entanglement, $I(T)\neq 0$ declaring that the bipartite is
correlated. The behavior of the  $C(\rho(T))$ and $I(T)$ are quite
different. For $I(T)$, at $T=0$ , $I(T)$ exhibits its maximum
value as given by (\ref{gr6new}). As the interaction time
increases, $I(T)$ goes down rapidly until local minima are
achieved, which result from the property of the modulus $"|...|"$
involved in the definition of the $I_{i,j}$. After that $I(T)$
decreases asymptotically to zero, i.e. $I(T)$ cannot exhibit SDE.
It would be more convenient to find the interaction time after
which $I(T)=0$. By means of (\ref{ebr5new})-(\ref{gr5new}), one
can easily prove that the solution of this equation is
$\exp(-\Gamma t)=0 $. This means that $I(T)$ needs infinite time
to vanish.
\begin{figure}[h]%
 \centering
 \subfigure[]{\includegraphics[width=8cm]{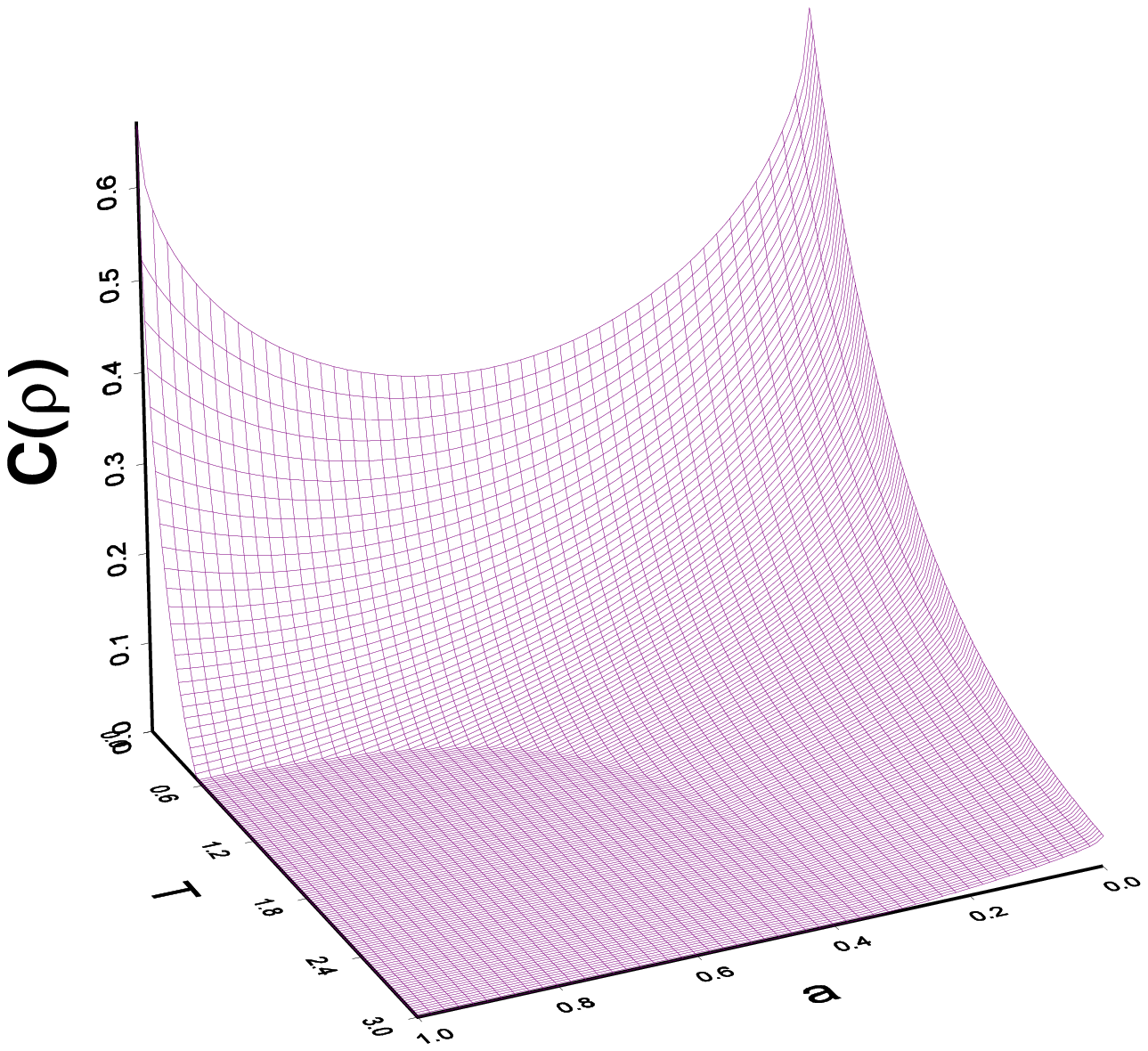}}
\subfigure[]{\includegraphics[width=8cm]{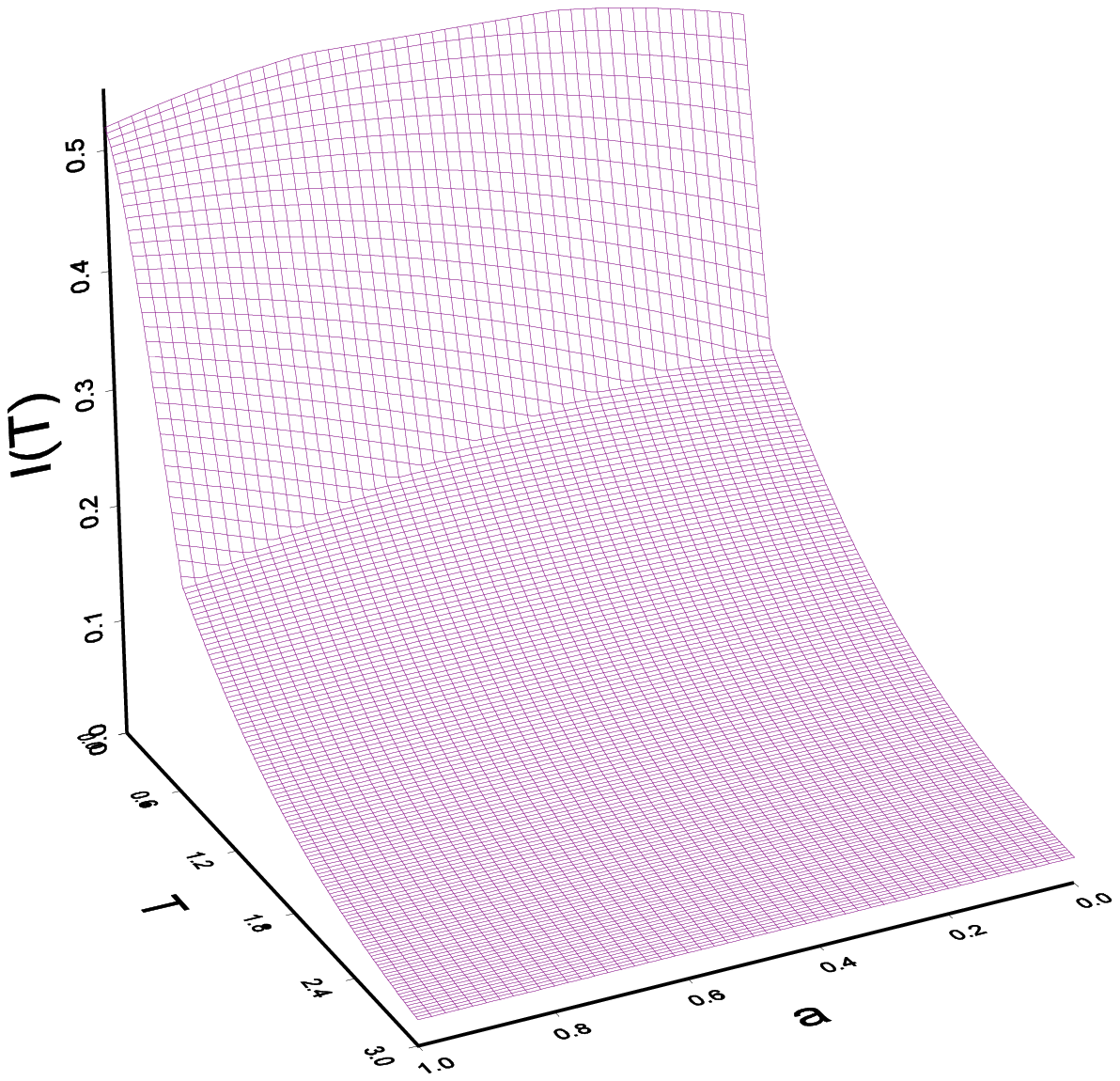}}
    \caption{Atom-atom entanglement  quantified by $C(\rho)$ (a) and
   $I$ (b)  when the atoms are initially prepared  in the mixed entangled state with
$b_{11}=a/3 \geq 0, b_{22}= b_{33}= b_{23}=b_{32}=1/3,\quad
b_{44}=(1-a)/3$.
   }
 \label{fig1}
\end{figure}
From this example and  the above one, one can see that it is
impossible to find sudden death of correlation/entanglement in the
framework of the $I$. The  reason for this is obvious: $I$ carries
information on both classical and nonclassical correlation. More
illustratively, SDE occurs when the entanglement evolves abruptly
to the classical one, however, the latter is still noticeable in
$I$. It is worthwhile mentioning that this behavior, the
disappearance of SDE, has already been addressed also via quantum
discord for dissipative two-qubit dynamics within Markovian and
non-Markovian environments \cite{wer1,wer2}. Choosing initial
conditions that manifest  SDE, the authors there have compared
  the dynamics of entanglement with that of   quantum discord.
The analysis
 showed that in all cases where the entanglement,  in terms of the concurrence,
 suddenly disappears  during a finite time interval,
the quantum discord vanishes only in the asymptotic limit or at
discrete instants. This reveals that  quantum discord is more
robust than the concurrence against SDE.

It is imperative  to extend the definition of the bipartite
indicator  $I$   to  multipartite systems. Indeed, in the
multipartite case, apart from fully separable and fully
correlated/entangled states, there also exists the notion of
partial separability. Here, we focus our discussion on  full
correlation/entanglement, however, the partial one can be
 treated similarly. For  multipartite systems, one has to deal
with all permutations of the correlation elements of the spin
operators. For instance, for the tripartite, one should consider
the correlation elements  of both two particles  and three
particles. However, this cannot be done so easily.
 Alternatively,
one may quantify the correlation for each two qubits individually
using (\ref{cor3}) and eventually carry  out the average over all
indicators. For fully correlated/entangled states, this has to be
done under the condition that the correlation must be available in
each pair, i.e. none of the individual indicators equals zero. In
this approach one usually exploits  the pairwise symmetry of
multipartite system, which
  simplifies the calculation of these indicators
considerably. In order to verify  this technique, it is sufficient
to compute  the correlation  of the three particle GHZ
(\ref{gr1}), which reads:
\begin{equation}\label{gr6}
  I=\frac{1}{3}[I^{(1,2)}+I^{(2,3)}+I^{(1,3)}]=1,
\end{equation}
where the indicator  $I^{(i,j)}$ denotes the correlation  between
particle $i$ and particle $j$, and we have used the results of the
example (1).
 The result given by (\ref{gr6})  represents nonclassical correlation and this agrees with the fact that
the GHZ is a maximally entangled tripartite state. Nonetheless,
this result has  already been deduced in terms of the $3$-tangle
$\tau_3$ (or residual tangle) \cite{wotter}, however, $\tau_3$
vanishes for a particular class of entangled three-qubit states
\cite{cerf}. This class is represented by the state $|W\rangle$:
\begin{equation}\label{cerf1}
    |W\rangle=\frac{1}{\sqrt{3}}[|e_1,g_2,g_3\rangle+|g_1,e_2,g_3\rangle+|g_1,g_2,e_3\rangle
    ].
\end{equation}
By applying  our criterion to (\ref{cerf1}),  we obtain $I=5/9$,
i.e.
 the $W$ state is entangled but not maximally. This shows that  the GHZ and
the $W$ state are locally inequivalent.

In conclusion, we have developed  a new criterion for quantifying
 correlation and/or entanglement of the  bipartite systems.
The proposed measure is simple and involves only calculation of
average values of the spin matrices, without the need for
operations on the density matrix of the system nor the calculation
of eigenvalues, as it happens with the concurrence.
  Also we have explained  how it
 can be extended to treat the correlation of  multipartite systems.
The criterion is based on observable quantities, which have a
clear physical meaning. This  can give  reliable physical
interpretation for  entanglement. Also  it suggests a way for
measuring the correlation  experimentally.
 We have  compared the results  obtained
from  our criterion with those from  the concurrence, and for
particular examples with the degree of separability. There is
agreement and disagreement between the behaviors of both criteria.
Generally, there is agreement  among the criteria in the
entanglement domain. The $C(\rho)$ and $I$ could have similar
behavior when $C(\rho)$ behaves smoothly. Moreover, from the
comparison with the other criteria, we have found, when $I>1/3$,
the bipartite is entangled, however, for $I\leq 1/3$ the bipartite
may be correlated classically or nonclassically.  Also, we have
shown that the indicator $I$ cannot exhibit  SDE. In this regard,
the behavior of $I$ is quite similar to that of  quantum discord
\cite{wer1,wer2}. Now, the question: is there any relationship
between  $I$ and  quantum discord? This point will be studied in a
future work where more general Hamiltonians and initial conditions
will be considered. From the analysis given in this Letter one can
arrive at the following fact: if it is proved that a physical
system is entangled in terms of any  criterion, it will
 be automatically   entangled in the framework of   $I$, i.e. $I\neq 0$, however, the reverse is not
true. Finally, we believe that the criterion given in this Letter
may have a wide range of potential applications in
 quantum information theory. For instance, the experimentalists
 always measure the  correlation in the quantum systems regardless  it is classical or nonclassical
 , since they cannot distinguish   between
 them. In this sense,
 the new criterion could be of interest.

\section*{ Acknowledgment}
 I would like to thank Professor F. F. Fanchini for drawing my
attention to the references [9,10],
 in which  SDE is not detected by the quantum discord.

\end{document}